\documentstyle[graphicx,12pt]{article}
\begin{document}

\small
\hoffset=-1truecm
\voffset=-2truecm
\title{\bf The instability of a black hole with $f(R)$ global monopole
under extended uncertainty principle}
\author {Hongbo Cheng\footnote {E-mail address:
hbcheng@ecust.edu.cn}\hspace {1cm}Yue Zhong\\
Department of Physics, East China University of Science and
Technology,\\ Shanghai 200237, China}

\date{}
\maketitle

\begin{abstract}
We consider the evolution of black hole involving an $f(R)$ global
monopole based on the Extended Uncertainty Principle (EUP). The
black hole evolutions refer to the instability due to the
Parikh-Kraus-Wilczeck tunneling radiation or fragmentation. It is
found that the EUP corrections make the entropy difference larger
to encourage the black hole to radiate more greatly. We also show
that the appearance of the EUP effects result in the black hole's
division. The influence from global monopole and the revision of
general relativity can also adjust the black hole evolution
simultaneously, but can not change the final result that the black
hole will not be stable because of the EUP's effects.
\end{abstract}

\vspace{6cm} \hspace{1cm} PACS number(s): 03.65.Bz, 03.65.Ta,
04.60.Bc\\
\vspace{4cm} \hspace{1cm} Keywords: EUP, $f(R)$ theory, global
monopole black hole, uncertainty principle

\newpage

\noindent \textbf{I.\hspace{0.4cm}Introduction}

According to the fact of accelerated expansion of the universe,
Buchdahl proposed the $f(R)$ theory as a kind of modified gravity
[1] and further the theory has been applied to explain the
accelerated-inflation problem without dark matter or dark energy
[2-4]. The $f(R)$ gravity generalizes the general relativity and
the generalization certainly arises in the description of the
background around the gravitational sources. The universe evolves
with decreasing temperature. In the process of the vacuum phase
transition in the early stage of the universe, several types of
topological defects such as domain walls, cosmic strings and
monopoles may have arisen [5, 6]. These topological defects formed
in favour of a breakdown of local or global gauge symmetries [7].
For example, a global monopole as a spherically symmetric
topological defect appears in the phase transition of a system
composed of a self-coupling triplet of a scalar field whose
original global $O(3)$ symmetry is spontaneously broken to $U(1)$
[6, 7]. It was shown that the structure of the metric outside a
monopole has a solid angle leading all light rays to be deflected
at the same angle [8]. In the case of a massive source involving a
global monopole in the universe with accelerated expansion, its
metric with terms associated with the monopole and $f(R)$ issue
are necessary [9, 10]. The nonvanishing modified parameter
$\psi_{0}$ from $f(R)$ theory belonging to the metric components
bring a cosmological horizon as a boundary of the universe to the
spacetime limited by the $f(R)$ global monopole [9]. More efforts
have been contributed to the model. it was found that the
parameter subject to the modification of gravity provides stable
circular orbits for massive test particles in the gravitational
field of an $f(R)$ global monopole [9, 10]. The quasinormal modes
for this kind of black holes were calculated with WKB
approximation [11-13]. The thermodynamics of the black hole with a
global monopole within the frame of $f(R)$ gravity was
investigated [14, 15]. The strong gravitational lensing for the
same models was discussed analytically [16]. The corrections from
the global monopole and the gravity modification in the $f(R)$
theory to the dominant term in the scattering absorption cross
section were computed in view of low frequency and small angles
[17].

In the research on black holes, there are more significant
measurement and theoretical predictions recently. One key
experiment is the Event Horizon Telescope (EHT) [18]. The EHT data
show the gravitational physics at the event horizon where no light
escapes from, which opens a window to probe the details of the
black hole core. The regions surrounding black holes are known as
the black hole shadows [18]. The black holes could be
fundamentally quantum objects regardless of their size, so the
quantum gravity effects can not be neglected for the physics of
microscopic black holes such as their evaporation profile and
singularity removal [19]. The quantum characteristics relate to
the horizon of black hole like the metric fluctuations [20-23] and
the quantum structures around the black hole [24-26]. It should be
pointed out that the quantum effects have something to do with the
Uncertainty Relation [24-26]. It is impossible to omit the
gravitational influence, so the terms with the Newtonian constant
would appear in the Heisenberg uncertainty principle [27-33].
These terms can be functions of momentum difference or distance
interval. The General Uncertainty Principle (GUP) with a series of
terms based on momentum difference is shown as a quantum
gravitational correction to the standard Heisenberg relation [28,
34]. One of its simple forms is chosen as,

\begin{equation}
\Delta x\Delta p\geq1+\beta l_{pl}^{2}\Delta p^{2}
\end{equation}

\noindent where $\beta$ is a constant of order unit and the
natural units $\hbar=c=1$ are utilized [28, 34]. Within the tiny
region, the momentum difference is large, so the deviations are
obvious according to the inequality (1). The GUP is used to study
the quantum gravity phenomenology of black holes and to cure the
divergence from states density near the black hole horizon while
relating the entropy of black hole to a minimal length as quantum
gravity scale [35-38]. The GUP also modifies the black hole
horizon and further changes the black hole entropy [35-38]. In an
anti-de Sitter spacetime, the Heisenberg uncertainty principle
should be deformed with a suitably chosen parameterization [39].
The Extended Uncertainty Principle (EUP) is shown as a
position-uncertainty correction to the Heisenberg inequality [28,
39]. We can select [28, 39],

\begin{equation}
\Delta x\Delta p\geq1+\alpha\frac{\Delta x^{2}}{L_{\ast}^{2}}
\end{equation}

\noindent where $\alpha$ is a constant of order unit and
$L_{\ast}$ is thought as a large fundamental distance scale. The
additional terms involve the ratio of distance difference and
distance scale according to Eq. (2). it is manifest that the EUP
introduces the quantum effects over the macroscopic distances [28,
39]. The EUP will redefine the horizon to revise the entropy [19].

Although the black holes are perceived as perfect absorbers
classically, their evolutions including the tunneling radiation
and fragmentation depending on the quantum mechanics and
thermodynamics respectively [40-47]. The tunneling formalism for
black holes subject to the imaginary part of action for
classically forbidden region of emission across the horizon is of
great concern [48-54]. With the help of the semi-classical
tunneling put forward by Kraus et.al., a lot of efforts have been
paid to the Hawking radiation of many kinds of objects such as BTZ
black holes [55-58], Taub-NUT black holes [59], Kerr-Newman black
holes [60-62], Godel black holes [63], etc.. The fragmentation of
black holes as a kind of evolution has also attracted more
attentions [47]. This thermodynamic instability was discussed
under non-perturbation [47]. The fragmentation issue was used to
probe the final fates of a series of black holes like the rotating
anti-de Sitter black holes [64], black holes with a Gauss-Bonnet
term [65] and charged anti-de Sitter black holes [66]. It should
be pointed out that the black holes tunneling radiation and
fragmentation both have something to do with their entropy
associated with their horizons [40-42]. The standard Heisenberg
uncertainty principle governs the horizons [27-34, 39]. As
mentioned above that the generalizations of the principle
undoubtedly modify the horizons and further the entropy, the GUP
has influence over the tunneling radiation and fragmentation of
black holes [28, 34-36, 39]. In the context of GUP modifying the
quantum mechanics [27-34, 67-72], the authors of Ref. [35, 36]
derived and estimate the relation between the Hawking tunneling
radiation of black holes and a minimal length as quantum gravity
scale in the higher dimensional spacetime by means of the
tunneling formalism. Under the GUP, we calculated the
Parikh-Kraus-Wilczeck tunneling radiation of black hole involving
an $f(R)$ global monopole to show that the square of momentum
difference term advances the emission of this kind of black holes
while the global monopole and the revision of general relativity
both hinder the black hole from emitting the photons [73]. We also
discovered that the same black hole keeps stable instead of
splitting without the GUP corrections [74]. Having researched on
the fragmentation of the black holes with $f(R)$ global monopole
in virtue of the second law of thermodynamics, we showed that the
influence from GUP leads the black hole to break into two parts
with larger mass and smaller ones respectively [74].

It is necessary to consider the tunneling radiation and
fragmentation of a Schwarzschild black hole with global monopole
under EUP within the frame of $f(R)$ scheme. As a kind of
generalization of Heisenberg uncertainty principle, the EUP is
significant and its quantum effects could appear at extremely
large scale [19, 28, 29]. The EUP revises the relation between the
horizon and the mass of black hole to correct the matter orbits
and innermost stable circular orbits (ISCO), size of the
photosphere [19]. The contribution of EUP corresponds to the dark
matter effects because the EUP correction fits the Milky Way's
rotation curve [19]. The thermodynamic properties of the
Schwarzschild black hole and the modified Unruh effect governed by
the EUP were discussed [75]. There must exist the massive objects
containing global monopoles in the spacetime with description of
$f(R)$ gravity as mentioned above. The EUP brings about the
effects on the tunneling radiation and fragmentation of black
holes through the EUP-corrected horizon [28, 34-36, 39]. During
our research on the evolution of the black holes, we should not
omit the corrections from EUP. To the best of our knowledge, few
efforts have been made to the investigation of the EUP influence
on the black hole stability due to the radiation and the
fragmentation. Under the EUP, we are going to derive and calculate
the entropy of a black hole swallowing $f(R)$ global monopole to
discuss the possibilities that the black hole radiates and breaks
into two sections with the help of the techniques of Ref. [43-46]
and Ref. [47] respectively. We wonder how the EUP affects the
possibilities. We list our results and compare the results with
those under the GUP finally.

\vspace{0.8cm} \noindent \textbf{II.\hspace{0.4cm}The tunneling
radiation of a black hole with an $f(R)$ global monopole under
extended uncertainty principle (EUP)}

We are going to investigate the entropy of a black hole with a
global monopole in the $f(R)$ theory. Under the corrections from
$f(R)$ statement, the spherically symmetric solution to the
gravitational field equation coupled to the matter field with a
spontaneously broken $O(3)$ symmetry was found [9, 10, 15],

\begin{equation}
ds^{2}=A(r)dt^{2}-B(r)dr^{2}-r^{2}(d\theta^{2}+\sin^{2}\theta
d\varphi^{2})
\end{equation}

\noindent where

\begin{equation}
A(r)=B^{-1}(r)=1-8\pi G\eta^{2}-\frac{2GM}{r}-\psi_{0}r
\end{equation}

\noindent and $G$ is the Newton constant. As a monopole parameter
in a typical grand unified theory, $\eta$ is of order $10^{18}GeV$
to lead $8\pi G\eta^{2}\approx10^{-5}$ [6, 8]. $M$ is mass
parameter. The factor $\psi_{0}$ represents the extension of the
standard general relativity. The roots of the equation $A(r)=0$
from metric (3) are [9, 10, 15],

\begin{equation}
r_{\pm}=\frac{1-8\pi G\eta^{2}\pm\sqrt{(1-8\pi
G\eta^{2})^{2}-8GM\psi_{0}}}{\psi_{0}}
\end{equation}

\noindent Here $r_{+}$ and $r_{-}$ stand for the outer radius and
inner ones respectively. It is obvious that the outer horizon will
disappear if the modified parameter $\psi_{0}$ vanishes.

The entropy of a black hole has something to do with the horizon
[40-43]. The corrected horizon radius certainly generalize the
expression of the entropy [19, 35, 36, 74, 75]. According to the
scheme of Ref. [19], the distance interval can be estimated as,

\begin{equation}
\Delta x'=\frac{\Delta x}{1+\frac{\alpha}{L_{\ast}^{2}}\Delta
x^{2}}
\end{equation}

\noindent where $\Delta x$ is the original size of black hole. If
the influence from EUP disappears like $\alpha=0$, the distance
difference $\Delta x'$ will recover to be original ones. The
$\alpha$-term from EUP shortens the black hole size according to
Eq. (6). Based on the approaches of Ref. [40-42], the Hawking
temperature for the black hole with the descriptions (3) and (4)
is a function of variables like $\eta$ and $\psi_{0}$,

\begin{equation}
T_{H}=\frac{1}{2\pi}(\frac{1-8\pi G\eta^{2}}{\Delta x}-\psi_{0})
\end{equation}

\noindent Here the interval can be let $\Delta x=2r_{H}$ and
$r_{H}=r_{-}$ is the black hole horizon without the extension of
Heisenberg's relation. The Bekenstein-Hawking entropy may be
obtained from the Hawking temperature (7) with the help of the
following thermodynamic relation [40-42, 46],

\begin{equation}
T_{H}=\frac{dE}{dS}\approx\frac{dM}{dS}
\end{equation}

\noindent In the emission process [46], the comparison between the
initial and the final values of the entropy of the black hole with
the solid deficit angle and $f(R)$ correction can be approximated
as,

\begin{equation}
\Delta S\approx-\frac{4\pi G}{(1-8\pi
G\eta^{2})^{2}}[M^{2}-(M-\hbar\omega)^{2}]-\frac{16\pi
G^{2}\psi_{0}}{(1-8\pi G\eta^{2})^{4}}[M^{3}-(M-\hbar\omega)^{3}]
\end{equation}

\noindent where $\omega$ is a shell of energy moving along the
geodesics towards the black hole with metric (3) [46, 56]. The
black hole's tunneling probability can be expressed as [46, 56],

\begin{equation}
\Gamma\sim e^{\Delta S}
\end{equation}

\noindent From Eq. (9), the higher order of typical grand unified
theory like increasing $8\pi G\eta^{2}$ will lead larger absolute
value of negative entropy difference, so will the farther away
from the Einstein's general relativity like increasing the
magnitude of $\psi_{0}$. It can be argued that the existence of
global monopole in the black hole or the deviation from general
relativity damps the emission of the black hole. The same topics
were considered under GUP as in Eq. (1). The emission of this kind
of black hole is promoted in favour of the greater parameter
$\beta$ as a coefficient of a quadratic term in the momentum
difference [73, 74].

It is significant to wonder how the EUP affects the entropy
difference associated with the tunneling probability of the black
hole described by the $f(R)$ global monopole metric. Following the
procedure of Ref. [40-42, 74], we choose the distance interval in
the temperature (7) as $\Delta x'$ shown in Eq. (6) to obtain the
EUP-corrected Hawking temperature,

\begin{eqnarray}
T'_{H}=\frac{1}{2\pi}(\frac{1-8\pi G\eta^{2}}{\Delta
x'}-\psi_{0})\nonumber\\
=T_{H}+\frac{\alpha}{\pi L_{\ast}^{2}}(1-8\pi G\eta^{2})r_{H}
\end{eqnarray}

\noindent while we also let $\Delta x=2r_{H}$ as above. The
corrections from EUP make the Hawking temperature higher. By means
of the thermodynamic relation (8) [40-42, 46], we derive the
corrected entropy difference of the radiating black hole involving
$f(R)$ global monopole as follow,

\begin{eqnarray}
\Delta S'=\Delta S'(\eta, \psi_{0}, \alpha)\hspace{6cm}\nonumber\\
=\frac{2\pi}{G}\int_{r_{H}}^{r'_{H}}\frac{-2\psi_{0}r_{H}^{2}+(1-8\pi
G\eta^{2})r_{H}}{\frac{4\alpha}{L_{\ast}^{2}}(1-8\pi
G\eta^{2})r_{H}^{2}-2\psi_{0}r_{H}+(1-8\pi G\eta^{2})}dr_{H}
\end{eqnarray}

\noindent Here $r_{H}=r_{-}$ and $r'_{H}=r_{-}|_{M\longrightarrow
M-\hbar\omega}$. It can be checked that before the performance of
integral $\Delta S'$ (12) will recover to be $\Delta S$ specified
by Eq. (9) if $\alpha=0$ [46]. Having performed the integral under
$\alpha\neq0$, we can show the change in entropy $\Delta S'$ in
unit of the difference for Schwarzschild black hole like $\Delta
S_{0}\approx-8\pi GM\hbar\omega$ [46],

\begin{equation}
\frac{\Delta S'}{\Delta
S_{0}}\approx-\frac{1}{\frac{\alpha}{L_{\ast}^{2}}}\frac{\psi_{0}}{4GM(1-8\pi
G\eta^{2})}\frac{1}{\sqrt{(1-8\pi G\eta^{2})^{2}-8GM\psi_{0}}}
\end{equation}

\noindent In the case of EUP, the tunneling probability of black
hole in Eq. (10) should be revised as $\Gamma'\sim e^{\Delta S'}$
[46]. The dependence of the entropy difference formulated for the
black hole including the $f(R)$ global monopole in Eq. (12) on the
variables $\alpha$ and $\psi_{0}$ corresponding to the EUP
correction and the deviation of standard gravity respectively is
plotted in Figure 1. The entropy change $\Delta S'$ is a
decreasing function of $\alpha$ for $\psi_{0}$ with a series of
definite values. The stronger influence from EUP leads the value
of $\Delta S'$ smaller, which retards the radiation of the black
hole, which is opposite to the case of GUP appearing in Eq. (1).
The considerable correction on the general relativity also causes
the black hole to be unstable due to the tunnel process.

\vspace{0.8cm} \noindent \textbf{III.\hspace{0.4cm}The
fragmentation instability of a black hole with an $f(R)$ global
monopole under extended uncertainty principle}

The fragmentation probability of a Schwarzschild black hole whose
spacetime has a solid deficit angle owing to a global monopole
dominated by $f(R)$ gravity should be discussed under EUP. We can
investigate the entropy of black hole to explore the its fate. In
view of Ref. [41, 77], the Bekenstein-Hawking entropy of black
hole is proportional to the horizon area,

\begin{equation}
S=\frac{1}{4}A_{H}
\end{equation}

\noindent where

\begin{equation}
A_{H}=4\pi r_{H}^{2}
\end{equation}

\noindent According to the second law of thermodynamics, the
thermodynamic argument for the fragmentation of a black hole
claimed that the black hole entropy must increase during its
evolution [47]. Here we assume that the black hole with $f(R)$
global monopole breaks into two parts involving the same kind of
monopole. In the process of fragmentation, the original black hole
can be thought as the initial state and the final state consists
of two black holes under the conservation of mass. Subject to the
second law of thermodynamics, we can compare the two entropies for
the initial and final state respectively to wonder whether the
fragmentation could happen. It was shown that the nature of the
entropy difference for the $f(R)$ global monopole black hole
limited by the Heisenberg inequality remains negative no matter
whether the general relativity has been generalized. The division
of this kind of isolated black holes can not occur spontaneously
[78]. When the GUP is introduced, the black hole containing the
$f(R)$ global monopole will split into two parts [78]. The
stronger influence from GUP can lead the difference of the masses
for the two fragmented black holes to be smaller [78]. As
mentioned above, the EUP is also a generalization of the
Heisenberg uncertainty principle. We should study the influence
from EUP on the fragmentation instability of the black hole
swallowing a global monopole governed by $f(R)$ theory. In the
initial case, the entropy of the isolated black hole can be
obtained from Eq. (14),

\begin{equation}
S_{i}=\pi r_{H}^{2}(M, \eta^{2}, \psi_{0})
\end{equation}

\noindent where $r_{H}(M, \eta^{2}, \psi_{0})=r_{-}$ shown in Eq.
(5). We estimate the black hole horizon amended by the EUP with
the original horizon like $\Delta x=2r_{H}(M, \eta^{2}, \psi_{0})$
from Eq. (6) [19],

\begin{equation}
r'_{H}(M, \eta^{2}, \psi_{0})=\frac{r_{H}(M, \eta^{2},
\psi_{0})}{1+\frac{4\alpha}{L_{\ast}^{2}}r_{H}^{2}(M, \eta^{2},
\psi_{0})}
\end{equation}

\noindent The correction will disappear with $\alpha=0$. Under
EUP, the corrected horizons of black holes lead to the corrected
entropy difference,

\begin{equation}
\Delta S'=S'_{f}-S'_{i}
\end{equation}

\noindent where

\begin{equation}
S'_{i}=\pi r'^{2}_{H}(M, \eta^{2}, \psi_{0})
\end{equation}

\begin{equation}
S'_{f}=\pi r'^{2}_{H}(\varepsilon_{M}M, \eta^{2}, \psi_{0})+\pi
r'^{2}_{H}((1-\varepsilon_{M})M, \eta^{2}, \psi_{0})
\end{equation}

\noindent By means of Eq. (17), we obtain the corrected radii
$r'_{H}(\varepsilon_{M}M, \eta^{2}, \psi_{0})$ and
$r'_{H}((1-\varepsilon_{M})M, \eta^{2}, \psi_{0})$ as follows,

\begin{equation}
r'_{H}(\varepsilon_{M}M, \eta^{2}, \psi_{0})=r'_{H}(M, \eta^{2},
\psi_{0})|_{M\longrightarrow\varepsilon_{M}M}
\end{equation}

\begin{equation}
r'_{H}((1-\varepsilon_{M})M, \eta^{2}, \psi_{0})=r'_{H}(M,
\eta^{2}, \psi_{0})|_{M\longrightarrow(1-\varepsilon_{M})M}
\end{equation}

\noindent It should be pointed out that $r'_{H}(M, \eta^{2},
\psi_{0})$ stands for the EUP-limited horizon of the initial black
hole swallowing the $f(R)$ global monopole. Under the EUP,
$r'_{H}(\varepsilon_{M}M, \eta^{2}, \psi_{0})$ and
$r'_{H}((1-\varepsilon_{M})M, \eta^{2}, \psi_{0})$ are the horizon
radii of the separated black holes belonging to the final state
with masses $\varepsilon_{M}M$ and $(1-\varepsilon_{M})M$
respectively. Following the same procedure as Ref. [66], we define
the mass distribution and its region $0\leq\varepsilon_{M}\leq1$.

It is fundamental to explore the fragmentation possibility of an
$f(R)$ global monopole black hole under the EUP. The sign of the
entropy difference during the evolution of the black hole helps us
to determine whether the black hole can split because the entropy
of a stable system cannot decrease in any spontaneous process [47,
79]. We ignore the generalization of the general relativity and
show the entropy difference in Eq. (18) as a function of the ratio
$\varepsilon_{M}$ graphically under EUP labelled as $\alpha$ in
Figure 2. We find that the entropy increases during the process
that the black hole becomes two new ones with
$0\leq\varepsilon_{M}\leq1$ and the existence of EUP. The
coefficient $\alpha$ can adjust the curves of entropy difference,
but it can not change the natures of the difference. The Figure 3
indicates that the deviation from the general relativity can also
adjust the entropy difference a little, but it can not let the
sign of the difference negative. The EUP encourages the global
monopole black hole under $f(R)$-generalized gravity to break into
two parts spontaneously, no limit to the distribution of the mass
of the black hole. The GUP also impels the same black hole to
break up, but weaker influence of the generalized principle will
cause the two new black holes to possess the different size, one
larger and the other smaller.

\vspace{0.8cm} \noindent \textbf{IV.\hspace{0.4cm}Discussion and
Conclusion}

We derive and compute the entropy difference of a black hole with
a $f(R)$ global monopole under the Extended Uncertainty Principle
to investigate the black hole evolution such as the
Parikh-Kraus-Wilczeck tunneling radiation and fragmentation. The
EUP adds a position-uncertainty term to the Heisenberg uncertainty
principle to reflect the quantum corrections to gravity in the
large scale [28, 39]. The EUP corrects the horizon of black hole
to change the black hole entropy. In favour of the entropy
difference modified by EUP in the process of emitting photons, the
stronger influence from EUP weakens the tunneling radiation of the
black hole with the description of $f(R)$ gravity while the global
monopole exists in the compact object, which is different from the
case of GUP. Under the Heisenberg inequality, the $f(R)$ global
monopole black hole keeps stable instead of splitting. In the case
of black hole fragmentation, the appearance of modification from
EUP will cause the black hole to divide into two objects with
arbitrary distributions of black hole mass contrary to the case
under GUP.

\vspace{3cm}

\noindent\textbf{Acknowledgement}

This work is supported by NSFC No. 10875043.

\newpage
\begin{figure}
\setlength{\belowcaptionskip}{10pt} \centering
\includegraphics[width=15cm]{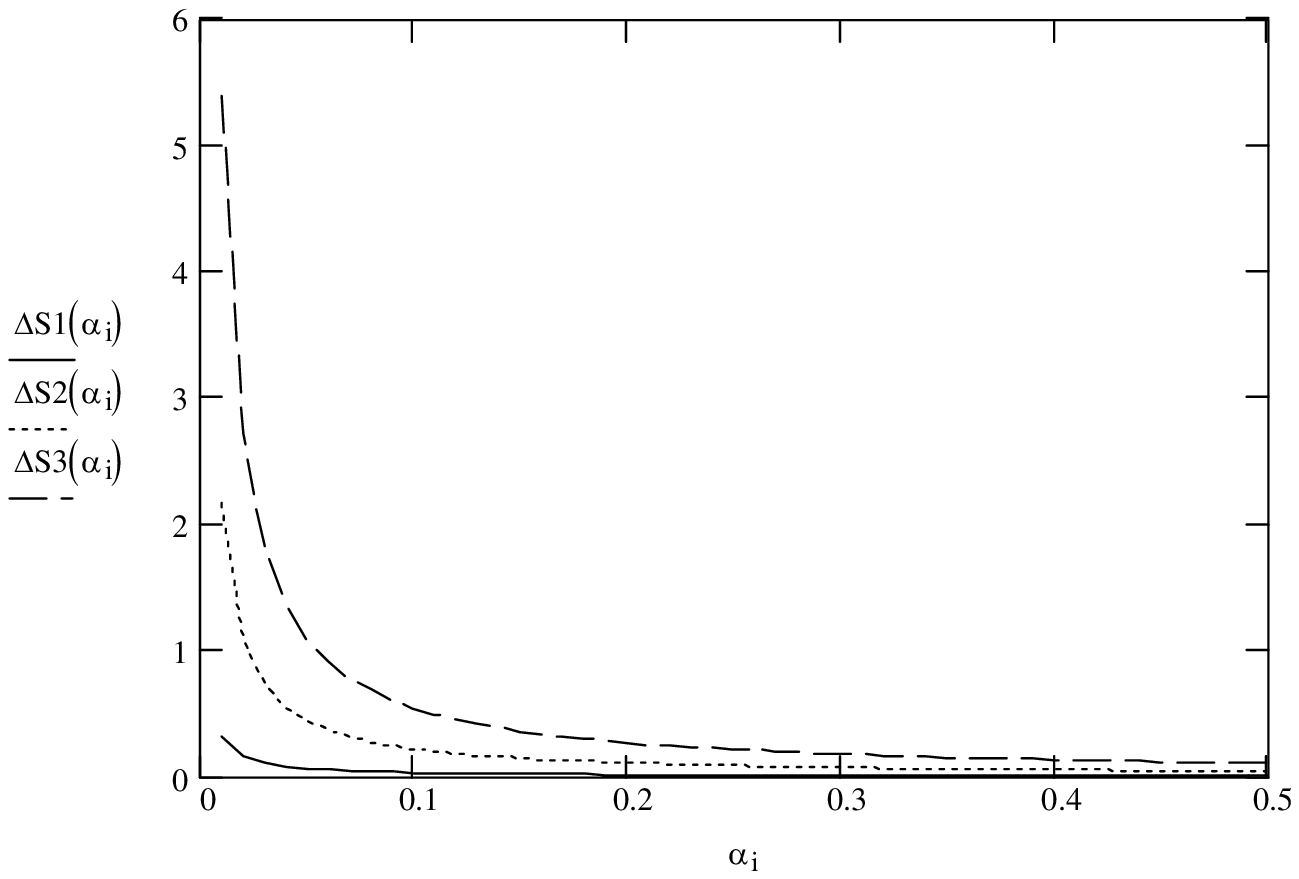}
\caption{The solid, dotted and dashed curves of the dependence of
entropy difference $\Delta S'$ on $\alpha$ for $\psi_{0}=0.01,
0.05, 0.08$ respectively and for simplicity $8\pi G\eta^{2}=0.1$
and $G=M=L_{\ast}=1$}
\end{figure}

\newpage
\begin{figure}
\setlength{\belowcaptionskip}{10pt} \centering
\includegraphics[width=15cm]{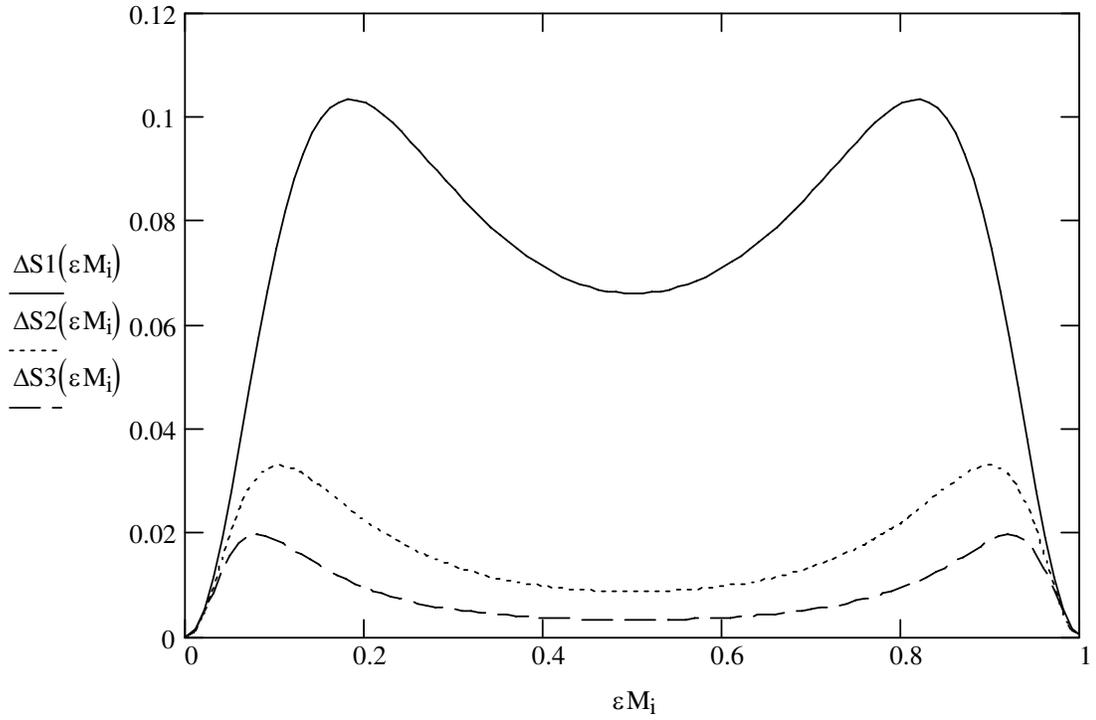}
\caption{The solid, dotted and dashed curves of the dependence of
entropy difference $\Delta S'$ on $\varepsilon_{M}$ for $\alpha=2,
6, 10$ respectively and for simplicity $8\pi
G\eta^{2}=0.1$,$\varepsilon_{\eta}=0.5$ and $G=M=L_{\ast}=1$}
\end{figure}

\newpage
\begin{figure}
\setlength{\belowcaptionskip}{10pt} \centering
\includegraphics[width=15cm]{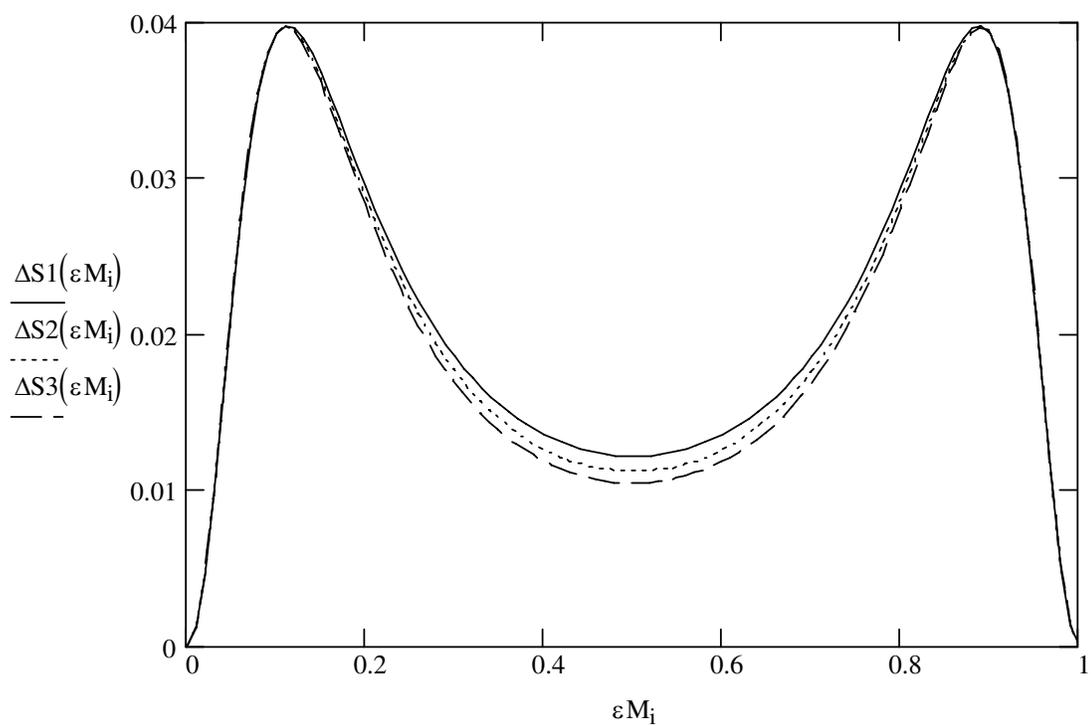}
\caption{The solid, dotted and dashed curves of the dependence of
entropy difference $\Delta S'$ on $\alpha$ for $\psi_{0}=0.01,
0.06, 0.1$ respectively and for simplicity $8\pi G\eta^{2}=0.1$,
$\alpha=5$, $\varepsilon_{\eta}=0.5$ and $G=M=L_{\ast}=1$}
\end{figure}

\end{document}